\documentclass[aip,apl,reprint,showpacs,superscriptaddress]{revtex4-1}

\usepackage{graphicx}
\usepackage{dcolumn}
\usepackage{bm}
\usepackage{amsmath,amssymb}
\usepackage{comment}

\def\U#1{{%
\def\O{\mbox{O}}
\def\u{\mbox{u}}
\mathcode`\u=\mu
\mathcode`\O=\Omega
\mathrm{#1}}}

\def\ii{{\mathrm{i}}}

\def\sub#1{_{\scriptsize\mbox{#1}}}

\begin{document}


\title{Frequency mixing at an electromagnetically induced transparency
like metasurface loaded with gas as a nonlinear element}

\author{Yasuhiro Tamayama}
\email{tamayama@vos.nagaokaut.ac.jp}
\author{Takuya Yoshimura}
\affiliation{Department of Electrical, Electronics and
Information Engineering, Nagaoka University of
Technology, 1603-1 Kamitomioka, Nagaoka, Niigata 940-2188, Japan}
\date{\today}

\begin{abstract} 

 Local electromagnetic field enhancement in resonant metamaterials is
 useful for efficient generation of nonlinear phenomena; however, the
 field enhancement is suppressed by losses of nonlinear elements in
 metamaterials. For overcoming this issue, we investigate the nonlinear
 response of an electromagnetically induced transparency like
 metasurface loaded with gas as the nonlinear element. To induce
 nonlinearity in the gas associated with discharges, an electromagnetic
 wave with a modulated amplitude is incident on the metasurface. The
 measured waveform and spectrum of the transmitted electromagnetic wave,
 along with light emission from the discharge microplasma, reveal that
 frequency mixing can occur on the metasurface. The parameter dependence
 of the conversion efficiency of the frequency mixing phenomenon shows
 that the efficiency is determined almost entirely by the ratio of the
 duration of microplasma generation to the modulation period of the
 incident wave amplitude. This result implies that the frequency mixing
 is derived from a binary change in the transmittance of the
 metasurface caused by the generation and quenching of the microplasma.
 
\end{abstract} 

\maketitle


There has been considerable interest in enhancing nonlinear phenomena
using metamaterials.\cite{lapine_14_rmp}
When an electromagnetic wave is incident on a
resonant metamaterial, the electromagnetic energy becomes compressed into a
small volume in the metamaterial. If a nonlinear element is excited by
the enhanced local electromagnetic field, nonlinear phenomena can be
efficiently generated. 
To date, various nonlinear phenomena, such as frequency
conversion,\cite{kanazawa_11_apl,nakanishi_12_apl,czaplicki_13_prl,celebrano_15_nat_nano,obrien_15_nat_mat,yang_15_nl,liu_16_nl,shorokhov_16_nl,alberti_16_apb,filonov_16_apl,gennaro_16_nl,metzger_16_apb,su_16_sci_rep,zhang_d_16_prb,wang_17_acsphoton,wolf_17_sci_rep,yang_17_acsphoton}
power-dependent
responses,\cite{liu_12_nat,ren_12_nat_com,fan_13_prl,guddala_16_ol,savinov_16_apl,keiser_17_apl,lawrence_18_nl}
and nonlinear wave
propagation\cite{shadrivov_06_photo_nano,noskov_12_prl,tamayama_13_prb,kozyrev_14_apl,agaoglou_14_bif,rai_15_prb}
have been investigated.

The enhancement factor of the local electromagnetic field
in a resonant metamaterial is inversely related with the losses in its unit structure.
In the optical region, nonlinear dielectric
resonators\cite{yang_15_nl,liu_16_nl,shorokhov_16_nl}
would be suitable for
the unit structure, since nonlinear dielectrics possess small losses.
In contrast, lower frequency regions, including microwave and terahertz,
require electronic devices, such as varactor diodes, to be introduced as
nonlinear elements.\cite{lapine_14_rmp}
These electronic devices have relatively high losses, which suppress the local electromagnetic field enhancement.

In this study, we investigate a metasurface loaded with gas as a
nonlinear element to develop a method for preventing the suppression of
the resonant enhancement of local electromagnetic fields.
Gases have low linear losses, and exhibit
discharges under the application of a strong electric field. Thus, we
consider gases to be suitable for the efficient generation of nonlinear
phenomena in resonant metamaterials, especially at lower
frequencies.
In addition, discharges cause abrupt changes in the electromagnetic
characteristics of gases. Therefore, it may be possible to generate highly
nonlinear phenomena in gas-loaded resonant metamaterials. 
In this paper, we show that frequency mixing can be generated using
the dynamic changes in the electromagnetic response of a metasurface associated
with a gas discharge in its unit structure.

\begin{figure}[tb]
 \begin{center}
  \includegraphics[scale=0.5]{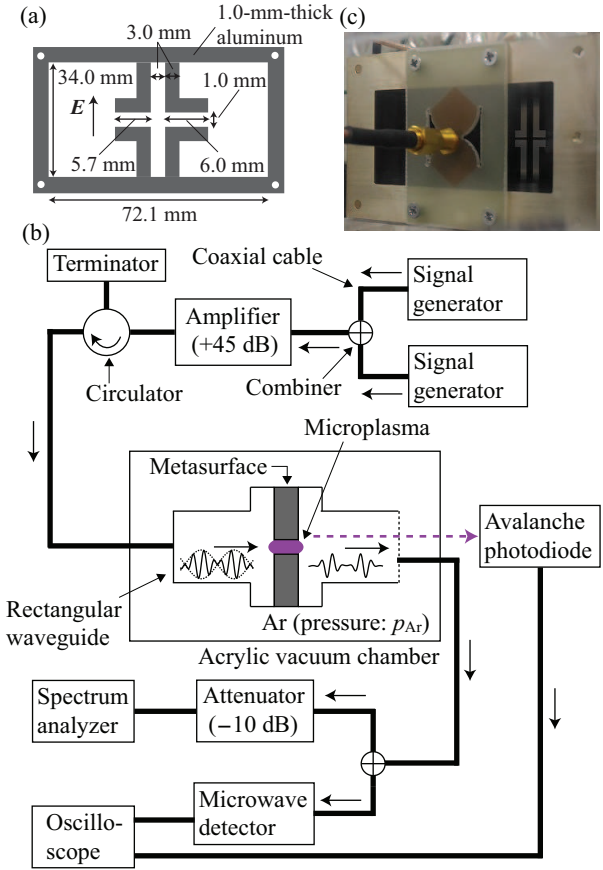}
  \caption{(a) Schematic of the structure of the EIT-like metasurface and (b)
  experimental setup. The metasurface structure was combined with the flange
  of a rectangular waveguide to place the metasurface at the center of
  the waveguide. Note that the walls of the waveguide acted almost like periodic boundaries in this experiment.
  (c) Photograph of output end of the waveguide with antenna. Microplasma
  emission can be observed through the triangular apertures in the antenna substrate. }
  \label{fig:setup}
 \end{center}
\end{figure}

Figure \ref{fig:setup}(a) shows the metasurface structure used in this study.
The details of the electromagnetic response of the metasurface are
described in our previous papers,\cite{tamayama_15_prb,tamayama_17_jap}
and we briefly review them here. 
This metasurface is composed of two radiatively coupled cut-wire resonators with
slightly different resonance frequencies, and exhibits an
electromagnetically induced transparency
(EIT)\cite{fleischhauer_05_rmp,alzar_02_ajp}
-like electromagnetic response.\cite{tamayama_14_prb}
The EIT-like transmission peak frequency is
$f_0 = 3.031\,\U{GHz}$, which is approximately the average of the resonance
frequencies of the two cut-wire resonators, $3.016\,\U{GHz}$ and
$3.059\,\U{GHz}$. When an electromagnetic wave with a frequency around $f_0$
is incident on the metasurface, a strongly enhanced local electric field
is induced at the gaps of the cut-wire resonators. The amplitude of the
local electric field at the gaps is about 300 times larger than that of the incident
field.
Because of the strong enhancement of the local electric field,
an electromagnetic wave with relatively low power can generate
a discharge microplasma at the gap of the cut-wire resonator with higher (lower)
resonance frequency when the incident frequency is larger (smaller)
than $f_0$.
After the generation of a microplasma at either of the gaps,
the electric field enhancement factor at the other gap drops to one
eleventh of its original value, because the microplasma itself behaves as a lossy
dielectric and changes the characteristics of the metasurface. 
Although microplasmas could be generated at both gaps simultaneously
by using higher incident power, high-powered electromagnetic waves were
not used in study and a microplasma was generated only at one gap.  

We consider a condition in which the EIT-like metasurface behaves as a
nonlinear medium. Harmonic generation was not previously observed on the
EIT-like metasurface with a microplasma, as described in our previous
paper.\cite{tamayama_15_prb} This implies that the metasurface with
a microplasma behaves as a linear medium for continuous waves.
The electromagnetic response of the metasurface must be dependent on the
incident field to induce a nonlinear phenomenon. That is, the generation
and quenching 
of the microplasma needs to be dependent on the incident field.
Therefore, we hypothesize that the metasurface would behave as a
nonlinear medium when exposed to electromagnetic waves with temporally
varying amplitudes.


We measured the transmission characteristics of the metasurface for
electromagnetic waves with sinusoidally modulated amplitudes to
investigate the nonlinear response of the metasurface. 
Figure \ref{fig:setup}(b) shows a schematic of the experimental setup in
this study.
The metasurface was placed in a rectangular waveguide with
cross-sectional dimensions of $34.0\,\U{mm} \times 72.1\,\U{mm}$.
The waveguide was placed in an acrylic vacuum chamber, where argon gas at a
pressure of $p\sub{Ar}$ was present. Two continuous waves with the same power were generated by two signal generators, and these waves were combined to generate a sinusoidally modulated
incident wave. The frequencies of these two continuous waves were
$f_1 = f_0 = 3.031\,\U{GHz}$ and $f_2 = f_1 + 10\,\U{kHz}$.
For this frequency condition, a microplasma was generated at
the gap of the cut-wire resonator with the higher resonance frequency, which is
at the left side of Fig.\,\ref{fig:setup}(a). 
The sinusoidally modulated wave was amplified using an amplifier. 
The power of each continuous wave after 
amplification was defined as $P_0$. 
The amplified wave was introduced
into the waveguide through a coaxial-to-waveguide transformer, and
was incident on the metasurface.
The transmitted wave was received by an ultra-wideband dipole
antenna.\cite{lule_05_mop}
The received wave was split into two
halves. One half was detected
by a microwave diode detector to observe the envelope of the
transmitted wave using an oscilloscope. The other half was fed
into a spectrum analyzer.
The microplasma emission at the gap of the cut-wire resonator with the higher
resonance frequency was observed through an aperture fabricated in the
receiving antenna substrate, as shown in Fig.\,\ref{fig:setup}(c), and detected by
an avalanche photodiode. The output signal of the photodiode was fed into the
oscilloscope.
We measured the envelope of the transmitted wave and the light
intensity of the microplasma emission, as well as the spectrum of the transmitted wave as
$P_0$ and $p\sub{Ar}$ were varied. 


\begin{figure}[tb]
 \begin{center}
  \includegraphics[scale=1.1]{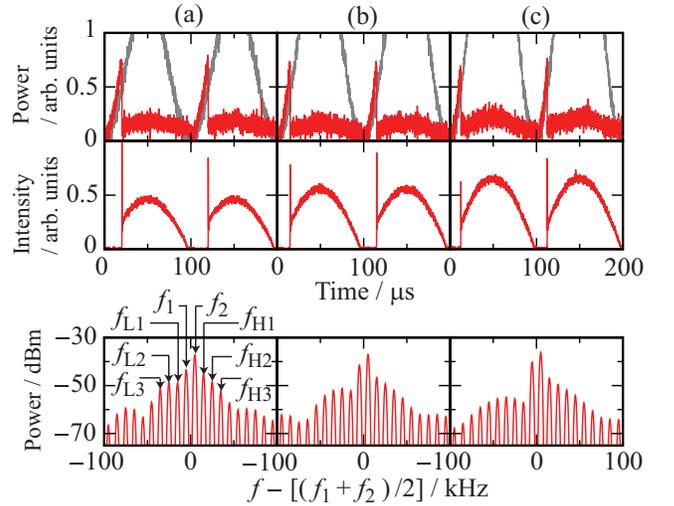}
  \caption{The envelope of the transmitted wave (upper row), 
  the light intensity of the microplasma emission (middle row), and the spectrum of the transmitted wave (lower row)
  at $p\sub{Ar}= 0.8\,\U{kPa}$ for (a) $P_0 = $ 29.1\,dBm, (b)
  30.6\,dBm, and (c) 32.1\,dBm. The light gray 
  curve in the upper panel shows the envelope of the transmitted wave
  when atmospheric-pressure air filled the chamber and a microplasma was not
  generated at the metasurface.}
  \label{fig:trans}
 \end{center}
\end{figure}

Figure \ref{fig:trans} shows examples of the measured characteristics of the transmitted
wave and microplasma emission.
The upper and middle panels show the envelope of the
transmitted wave and the intensity of the microplasma emission, respectively.
As the incident amplitude increases from zero,
the transmission amplitude first increases and then drops sharply.
Further increases and subsequent decreases in the incident amplitude cause little change in the
transmission amplitude. After the incident amplitude becomes smaller
than a certain value, the transmission amplitude varies with
the incident amplitude again.
The light intensity of the microplasma emission varies periodically, with a period equal
to the modulation period of the incident amplitude. 
The duration of the low-transmittance state agrees with the time for
microplasma emission. Thus, the temporal variation of the transmittance is confirmed
to be caused by the generation and quenching of the microplasma.
The result that the transmittance becomes low during 
microplasma generation agrees with our previous
results.\cite{tamayama_15_prb,tamayama_17_jap}
As $P_0$ increases, the ratio of the duration of microplasma generation
to the modulation period increases. This is because the duration that the
incident amplitude exceeds the threshold value for plasma ignition
increases with $P_0$. 
The lower panel of Fig.\,\ref{fig:trans} shows the spectrum of the
transmitted wave.
Many sidebands are observed in the spectrum and 
the power of each sideband varies with $P_0$.
The frequency separation
between the sidebands is $10\,\U{kHz}$, which equals $f_2 - f_1$.
These results in both the time and frequency domains
clarify that frequency mixing can be realized using the generation
and quenching of a microplasma on the EIT-like metasurface.

\begin{figure}[tb]
  \begin{center}
   \includegraphics[scale=0.6]{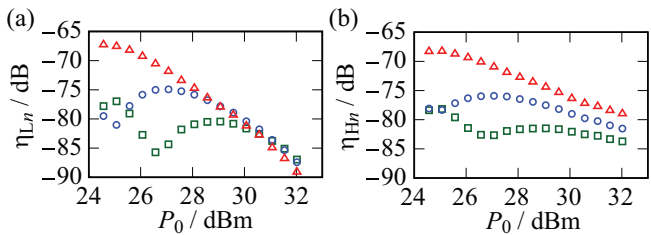}
   \caption{Incident power dependence of frequency conversion efficiency
   at $p\sub{Ar}=0.8\,\U{kPa}$ for (a)
   lower sidebands and (b) upper sidebands.
   Triangles, circles, and squares correspond to $n=$ 1, 2, and 3,
   respectively.}
  \label{fig:eff_power}
  \end{center}
\end{figure}

Next, we investigate the frequency mixing phenomenon in detail. 
Figure \ref{fig:eff_power} shows the incident power dependence of the
frequency conversion efficiency. Here,
we define the frequency conversion
efficiency as $\eta_{\{\mathrm{L,H}\}n} = P_{\{\mathrm{L,H}\}n} / P_0$
($n=1,2,3,\cdots$), where
$P_{\{\mathrm{L,H}\}n}$ is the power of a sideband
with frequency $f_{\{\mathrm{L,H}\}n}$, as indicated in
Fig.\,\ref{fig:trans}.
The conversion efficiency is increased or decreased by increasing the
incident power, depending on $P_0$ and $n$. 
This observation is quite different compared with typical nonlinear phenomena, in which
frequency conversion efficiency always increases with incident power.


\begin{figure}[tb]
 \begin{center}
  \includegraphics[scale=0.6]{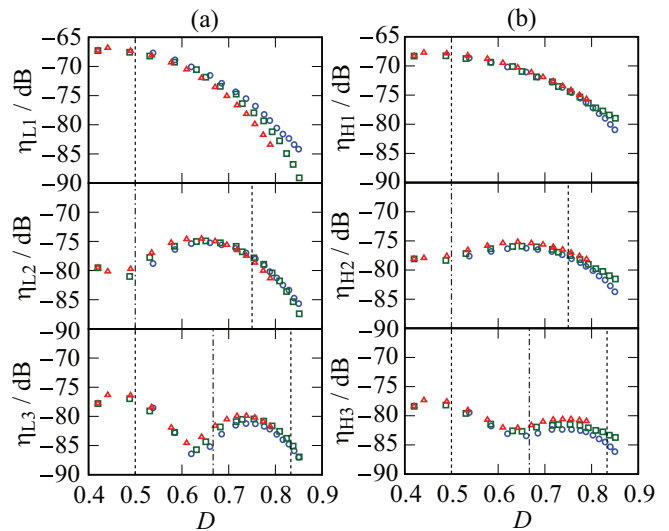}
  \caption{Dependence of frequency conversion efficiency for (a) lower
  sidebands and (b) upper sidebands on the ratio of plasma
  generation duration to modulation period.
  Triangles, squares, and circles correspond to experimental data at
  $p\sub{Ar} =$ 0.4\,kPa,
  0.8\,kPa, and 1.2\,kPa, respectively. Vertical dashed lines and
  dot-and-dash lines represent values of $D$ for which 
  $|c_n|$ becomes a local maximum or minimum,
  respectively.}
  \label{fig:eff_duty}
 \end{center}
\end{figure}

Let us consider a simple model
of the EIT-like metasurface to understand the dependence
of the frequency conversion efficiency on $P_0$.
The transmittance of the metasurface with a microplasma
is relatively low, and nearly independent of the incident
amplitude.
Therefore, we assume that the transmittance of the metasurface only
depends on whether the microplasma is generated or not. That is, the
amplitude transmittances with and without the microplasma are assumed to be
$a_1$ and $a_2$ ($a_2 > a_1$), respectively.
The time dependence of the amplitude transmittance $a(t)$ is a periodic
function with a period of $T= 100\,\U{us} = (f_2 - f_1)^{-1}$ in
this experiment. Thus, $a(t)$ can be expanded into Fourier series as
follows: 
$\displaystyle
a(t) = Da_1 + (1-D) a_2 + \sum_{n=1}^{\infty}
\left[c_n \exp{(-\ii 2n \pi t / T)} + c_n^* \exp{(\ii 2n \pi t / T)} \right]$,
where $D$ is the ratio of the duration of microplasma generation to $T$,
$\displaystyle
|c_n| = [(a_2 - a_1)/(n \pi)]|\sin{[n\pi (1-D)]}|$,
and * indicates the complex conjugate.
The time dependence of the transmitted amplitude is described by
the product of the incident amplitude and $a(t)$. Therefore, the spectrum of
the transmitted wave can be expressed as the convolution of the Fourier transforms of the
incident wave and $a(t)$. Assuming that the lower (upper) sidebands can be expressed
as the convolution of the Fourier transforms of a continuous wave with
frequency of $f_1$ ($f_2$) and $a(t)$ for simplicity, the powers of the 
sidebands are described by $|c_n|$, where the only parameter
that depends on the experimental condition is $D$. 

To compare this model with the experiment, we show
in Fig.\,\ref{fig:eff_duty} the dependence of $\eta_{\{\mathrm{L,H}\}n}$
($n=1,2,3$) on $D$ that was measured for various values of
$P_0$ ($\leq 32.1\,\U{dBm}$) at $p\sub{Ar} = 0.4\,\U{kPa}$, $0.8\,\U{kPa}$,
and $1.2\,\U{kPa}$. 
Here, $D$ was determined from the light intensity of the microplasma emission.
The conversion efficiencies are found to be determined almost entirely by
$D$, even though $P_0$ and $p\sub{Ar}$ are tested at various values. 
In addition, the values of $D$ for which $\eta_{\{\mathrm{L,H}\}n}$
becomes local maxima or minima in the experiment roughly agree with those
calculated from the model.
This implies that the frequency mixing is derived from the binary change
in the transmittance, which is caused by the generation and quenching of the microplasma.
The discrepancy between the experiment and model in $D$ for which $\eta_{\{\mathrm{L,H}\}n}$ becomes a local maximum or minimum
is relatively large in the region of larger $D$ values, which may be caused by oversimplifications in the model.

The nonlinear response in the present experiment cannot be
expressed as a $n$-th order nonlinearity.
In typical nonlinear media, 
the ratio of the polarization density to the electric field depends on the
electric field. However, the ratio in the metasurface
depends instead on the amplitude of the electric field. 
In this way, the nonlinear response at this metasurface is similar to
that of a rf-SQUID metamaterial.\cite{zhang_d_16_prb}
The abrupt change in the electromagnetic response
observed in these metamaterials
may be useful for the efficient generation of various nonlinear phenomena.


In conclusion, we investigated the nonlinear response of an EIT-like metasurface loaded
with gas as a nonlinear element to develop a method for preventing the
suppression of the enhancement of
the local electromagnetic field in a resonant metamaterial.
We measured the transmission characteristics of the metasurface and the light
intensity of the microplasma
emission for an incident wave with a sinusoidally modulated amplitude to
verify the generation of nonlinear phenomenon. 
The measured envelope and spectrum of the transmitted wave, as well as the 
temporal dependence of the microplasma emission, revealed that
frequency mixing occurred at the metasurface because of the generation and
quenching of the microplasma. The parameter dependence
of the frequency conversion efficiency showed that the efficiency was determined
almost entirely by the ratio of duration of the microplasma generation to the modulation
period.
The metasurface may be classified into a kind of waveform selective
metasurface\cite{wakatsuchi_15_jap,wakatsuchi_16_sci_rep}
because the time variation of the incident amplitude
is essential for the generation of nonlinear phenomenon.
Note that the nonlinear phenomenon in this study is not caused by a
nonlinearity in the plasma, in contrast with some previous
works.\cite{iwai_15_pre,iwai_15_apex,gregorio_16_psst}
Although harmonic generation has not been observed in this metasurface
with microplasma,
it can be generated in principle and should be
investigated in a future study. 
It should be added that the interactions between free electrons in
conductors and the enhanced local magnetic field 
can be used as another way to prevent the suppression of the enhancement of the local
electromagnetic field for efficient generation of nonlinear
phenomena.\cite{wen_17_prl}

This research was supported in part by \mbox{JSPS} \mbox{KAKENHI} Grant
Number \mbox{JP16K14249}, and by a research grant from The Murata
Science Foundation. 


%

\end{document}